\documentclass[prb,aps,twocolumn,floatfix]{revtex4}
\usepackage{textcomp}
\usepackage{graphicx}% Include figure files
\usepackage{amssymb}
\usepackage{epsfig}

\begin{document}

\title{Electrical transport in suspended and double gated trilayer graphene}

\author{Thymofiy Khodkov}
\author{Freddie Withers}
\author{David Christopher Hudson}
\author{Monica Felicia Craciun}
\author{Saverio Russo}
\affiliation{Centre for Graphene Science, College of Engineering, Mathematics and Physical Sciences, University of Exeter, Exeter EX4 4QF, UK}

\begin{abstract}

We present a fabrication process for high quality suspended and double gated trilayer graphene devices. The electrical transport measurements in these transistors reveal a high charge carrier mobility (higher than $20000 cm^2/Vs$) and ballistic electric transport on a scale larger than $200nm$. We report a particularly large on/off ratio of the current in ABC-stacked trilayers, up to 250 for an average electric displacement of -0.08 V/nm, compatible with an electric field induced energy gap. The high quality of these devices is also demonstrated by the appearance of quantum Hall plateaus at magnetic fields as low as $500mT$.
\end{abstract}

\maketitle

The unique combination of physical properties found in graphene materials \cite{CastroNeto2009} -one or few layers of carbon atoms on a honeycomb lattice- holds promise for future applications ranging from high frequency \cite{Avouris2011} to flexible and transparent electronics \cite{Samsung}. For example, few-layer graphene (FLG) are the only known materials to exhibit an electric field- and stacking-dependent band structure. While Bernal (AB-) stacked bilayers and rhombohedral (ABC-) stacked trilayers display an electric field tunable band-gap \cite{McCann2006,Oostinga2008,Crommie2009,Peeters2009,Koshino2010,Heinz2011,Jhang2011,Lau2011}, ABA-stacked trilayers are semimetals with an electric field tunable band overlap between conduction and valence bands \cite{Craciun2009,Koshino2009}.

Experimentally, the electric field control over the band structure of FLGs is readily obtained in double gated geometries \cite{Russo2009,Craciun11}. These devices comprise a FLG conductive channel embedded between a top and a bottom gate. The gate voltages applied to the two gates allow the independent control of the Fermi energy and of the perpendicular electric field applied onto the FLG. Most of the experimental work in double-gated structures has focussed on FLG in direct contact with the two oxide gate dielectrics \cite{Craciun11}, and only recently two independent studies have reported suspended double-gated bilayer graphene \cite{Martin2010,Velasco}, whereas no report has appeared yet on thicker few-layer graphene. Though easy to fabricate, the supported devices are difficult to anneal and suffer of the presence of charge traps in the oxide dielectric layer \cite{Craciun11}. Even though, the electron mobility in supported structures can be significantly improved by the use of BN top- and bottom-dielectric layers \cite{Mayo}, suspended devices offer the possibility to explore the coupling between electronic properties and mechanical vibrations. Therefore, these structures are suitable for investigating strain-induced band structure modifications \cite{Eros}.

\begin{figure}[ht]{}
\includegraphics[scale=0.2]{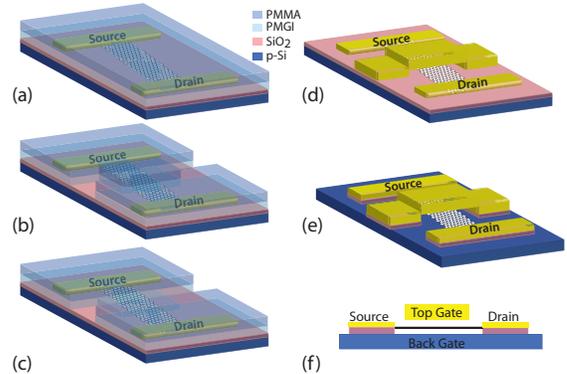}
\caption{\label{fig1} Schematic representation of the suspended and double gated graphene devices fabrication procedure.}
\end{figure}

Here we use a fabrication technique for suspended and double gated structures based on organic polymers and developers \cite{Lau2008}, which avoids any damage and/or contamination of the graphene caused by the deposition of a top-gate oxide dielectric (e.g. by Atomic Layer Deposition or by electron-beam evaporation) \cite{Russo2009,Craciun11,Martin2010}. The air-gap top gate allows the \textsl{in situ} annealing of the suspended FLG, necessary to improve the electrical performances of the devices \cite{Bachtold2007}. The processing described in this letter invariably delivers high mobility devices and ballistic electric transport on a length scale $> 200 nm$. Here we present electrical transport measurements in high quality supended double-gated trilayer devices. We report the appearance of well defined quantum Hall plateaus at magnetic fields as low as $500mT$. This technique can easily be extended to the fabrication of electrostatically defined quantum dots and to electron focussing experiments in double gated graphene materials.

Figure 1 illustrates a schematic of the fabrication process for suspended and double gated graphene devices. The starting device is a graphene transistor fabricated on SiO2 (300nm)/p-Si which serves as a back-gate. The processing of these multi-level structures consists of the following steps. At first, the regions of the pillars of the air-gap bridges are patterned in a PMGI/PMMA bilayer (100nm/600nm thick) by electron-beam (e-beam) lithography followed by development of both layers in MIBK and MF319, see Fig. \ref{fig1}a and b. The fabrication of the bridge beam \cite{Lau2008} involves the exposure and development of the PMMA layer only in MIBK -which leaves unaffected the PMGI underneath (Fig. \ref{fig1}c). The bridge fabrication is concluded by the evaporation and lift-off of a Cr/Au (20nm/180nm). The remaining PMGI sacrificial-layer is then removed by wet-etching in MF319 (Fig. \ref{fig1}d). This organic spin-on sacrificial layer does not induce structural defects and, it is easy to anneal any residual left by it. In contrast, the evaporation of SiO2 sacrificial layer used in previous experiments \cite{Martin2010} can create structural defects which irreversibly affect the electronic properties of graphene. The suspension of the graphene samples is then accomplished by standard wet-etching of $150 nm$ of $SiO_2$ in a solution of buffered HF (Fig. \ref{fig1}e and f).

\begin{figure}[ht]{}
\includegraphics[width=0.45\textwidth]{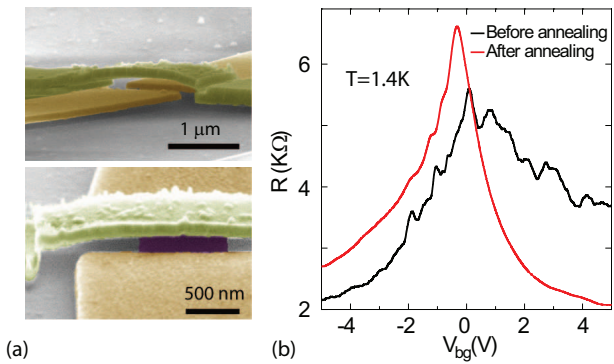}
\caption{\label{fig2} (a) False colour SEM micrograph of a suspended and double gated graphene device. (b) Resistance \textsl{versus} back-gate voltage ($V_{bg}$) before and after current annealing for a double gated trilayer graphene device.}
\end{figure}

Special care has to be taken during the final drying process of the sample since the surface tension of the liquids and the capillary forces can easily cause the collapse of the nano-structures. A common solution to this problem is to dry the samples in a critical point dryer (CPD) -making use of the zero surface tension in the supercritical transition of $CO_2$. However, after being dried in the CPD, the graphene surface is often covered by contaminants present in the liquids and/or the $CO_2$ gas used in the process. These contaminants dope the graphene, degrade its electrical properties such as the charge carrier mobility and they are also very difficult to anneal. Here we undertake an alternative route to dry the samples after etching, making use of the fact that both surface tension and capillary forces are temperature dependent -i.e. they decrease when approaching the boiling point of the liquids. Simply warming up the IPA at $50 ^\circ$C reduces significantly the surface tension of this liquid, making it possible to suspend the double-gated structures by just leaving them to dry in atmosphere. This procedure invariably delivers suspended double-gated graphene devices with flakes as large as $3 \mu m$ wide and up to $2 \mu m$ long. Fig. \ref{fig2}a shows a false colour Scanning electron microscope (SEM) micrograph of a typical suspended and double gated graphene device taken under a shallow angle to highlight the multi-level structure comprising the air-gap top-gate and the suspended flake.

\begin{figure}[ht]{}
\includegraphics[width=0.45\textwidth]{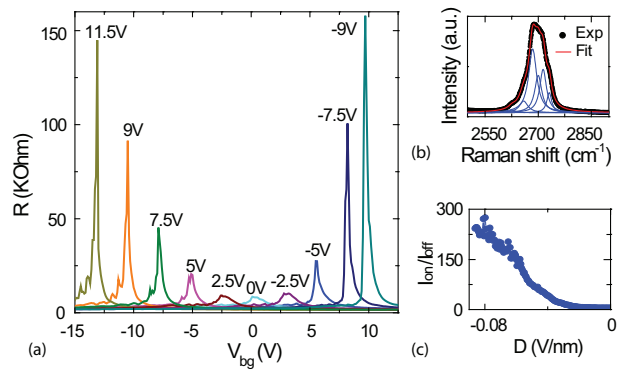}
\caption{\label{fig3} (a) Resistance \textit{vs.} back-gate voltage ($V_{bg}$) measured at $T=0.3K$ and for different values of fixed top-gate voltage ($V_{tg}$) as indicated in the graph. (b) 2D-Raman peak measured with a 532nmn laser, 5mW power and a spot size of 1.5 $\mu m$. The dots are the experimental data points, whereas the red continuous line is a fit to 6 Lorentzians (continuous blue lines). (c) Measurements of the on/off ratio of the current ($I_{on}/I_{off}$) as a function of the average electric displacement $D$.}
\end{figure}

We have characterized the electrical properties of these suspended and double-gated devices measuring the resistance with standard lock-in technique in a current- or voltage-biased configuration and in the linear regime -i.e. the excitation current (voltage) was varied to ensure that the voltage drop across the sample was smaller than the temperature broadening of the Fermi distribution. All the devices are current annealed \textsl{in situ} -i.e. in high vacuum ($10^{-6}$ mbar) and at low temperature $T=4$K with current densities as high as $1.4 mA/ \mu m^2$. Upon annealing the residual doping of the samples is reduced to zero and the charge carrier mobility typically increases by at least one order of magnitude, see Fig. \ref{fig2}b. In total we have studied more than 5 double gated FLG devices, and in this letter we discuss the representative data of an ABC-stacked trilayer graphene.

\begin{figure}[ht]{}
\includegraphics[width=0.35\textwidth]{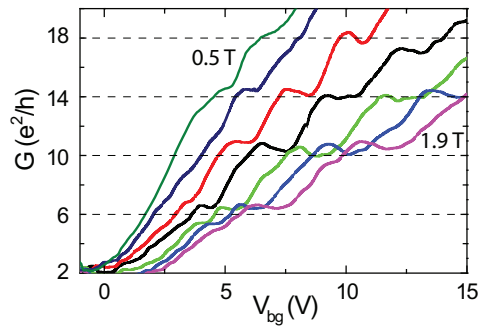}
\caption{\label{fig4} Conductance $\textit{vs.}$ back-gate voltage measured at $V_{tg=0V}$ for different values of perpendicular magnetic field from 0.5T up to 1.9T in steps of 0.2T.}
\end{figure}

Both the stacking-order and the number of layers were reliably identified by means of Raman spectroscopy as previously reported \cite{Malard2009,Jhang2011,Lau2011,LuiHeinz2011}. In particular the peak at $2700 cm^{-1}$ (2D-peak) in the Raman spectra of graphene depends on the band structure of the material. In trilayer graphene experimentally a minimum number of 6 Lorentzian functions can be used to describe the shape of the 2D peak, whereas the asymmetry of this peak (with a pronounced shoulder) identifies the rhombohedral stacking order (see Fig. \ref{fig3}b). Fig. \ref{fig3}a shows the 2-terminal resistance measured at T=0.3K as a function of back-gate voltage ($V_{bg}$) for different fixed values of top-gate voltage ($V_{tg}$). It is apparent that the maximum of resistance increases with increasing the external perpendicular electric field. This observation is consistent with the opening of an electric field induced band gap in the energy dispersion of rhombohedral trilayer graphene \cite{Peeters2009,Koshino2010,Heinz2011,Jhang2011,Lau2011}. The high quality of these samples is demonstrated by the fact that we observe a particularly high on/off ratio of the current ($I_{on}/I_{off}$). If we define the average electric displacement as $D=(D_{bg}+D_{tg})/2$ with $D_{bg}=\frac{\varepsilon}{d(\varepsilon+1)}V_{bg}$ and $D_{tg}=V_{tg}/d$ ($d=150nm$ and $\varepsilon=3.9$ for $SiO_{2}$), we find that Ion/Ioff equals 250 for $D=-0.08 V/nm$. The Ion/Ioff value typically found in our devices is at least twice as large as previously reported in supported double gated bilayer graphene devices \cite{Avouris2010}.

Electrical transport measurements in perpendicular magnetic field reveal the formation of Landau levels (LLs) starting from 0.5T, with the appearance of quantized plateaus in the conductance (see Fig. \ref{fig4}). Experimentally we find that the unique quantization sequence of the rhombohedral trilayers $G (e^2/h)= \pm 6, \pm 10, \pm 14, \pm 18 ...$ becomes clearly visible at magnetic fields as low as 0.9T. We identify the filling factors of the Quantum Hall plateaus at $\nu= 6, 10, 14, 18$ with $\nu = n_{s} \phi_{0}B^{-1}$, where $\phi_{0}$ is the flux quantum, $n_{s}$ is the charge carrier density calculated using the known capacitance to the back gate and B is the magnetic field. The observed plateaus are expected from the 3-fold degenerate zero-energy LLs of the ABC-stacked trilayer graphene ($E_{n}\propto B^{3/2}\sqrt{n(n-1)(n-2)}$) with 4-fold spin and valley degeneracy \cite{Guinea2006,Katsnelson2011,Kumar2011,Jhang2011,Zaliz2011}.

The high quality ABC-stacked trilayer devices invariably produced by this processing make it possible to access the quantum Hall physics at magnetic fields which are 2 orders of magnitude smaller than previously experimentally reported in supported double-gated devices \cite{Russo11}.
S.R. and M.F.C. acknowledge financial support from EPSRC (Grant No. EP/G036101/1 and no. EP/J000396/1) and from the Royal Society Research Grant 2010/R2 (Grant no. SH-05052).


\begin{thebibliography}{100}


\bibitem{CastroNeto2009} A. H. C. Neto, F. Guinea, N. M. R. Peres, K. S. Novoselov, and A. K. Geim, \textit{Rev. Mod. Phys.} \textbf{81}, 109 (2009).
        
\bibitem{Avouris2011} Y. Wu, Y. Lin, A. A. Bol, K. A. Jenkins, F. Xia, D. B.
Farmer, Y. Zhu, and P. Avouris, \textit{Nature} \textbf{472}, 74 (2011). 

\bibitem{Samsung} S. Bae, H. Kim, Y. Lee, X. Xu, J.-S. Park, Y. Zheng,
J. Balakrishnan, T. Lei, H. R. Kim, Y. I. Song, et al., \textit{Nat. Nano.} \textbf{5}, 574 (2010).

\bibitem{McCann2006} E. McCann, \textit{Phys. Rev. B} \textbf{74}, 161403 (2006).

\bibitem{Oostinga2008} J. B. Oostinga, H. B. Heersche, X. Liu, A. F. Morpurgo,
and L. M. K. Vandersypen, \textit{Nat. Mater.} \textbf{7}, 151 (2008).

\bibitem{Crommie2009} Y. Zhang, T. Tang, C. Girit, Z. Hao, M. C. Martin,
A. Zettl, M. F. Crommie, Y. R. Shen, and F.Wang, \textit{Nature}
\textbf{459}, 820 (2009).

\bibitem{Peeters2009} A. A. Avetisyan, B. Partoens, and F. M. Peeters, \textit{Phys. Rev. B} \textbf{80}, 195401 (2009).

\bibitem{Koshino2010} M. Koshino, \textit{Phys. Rev. B} \textbf{81}, 125304 (2010).

\bibitem{Heinz2011} C. H. Lui, Z. Li, K. F. Mak, E. Cappelluti, and T. F. Heinz,
\textit{Nat. Phys.} \textbf{7}, 944 (2011).

\bibitem{Jhang2011} S. H. Jhang, M. F. Craciun, S. Schmidmeier, S. Tokumitsu,
S. Russo, M. Yamamoto, Y. Skourski, J. Wosnitza,
S. Tarucha, J. Eroms, and C. Strunk, \textit{Phys. Rev. B} \textbf{84}, 161408 (2011).

\bibitem{Lau2011} W. Bao, L. Jing, J. Velasco, Y. Lee, G. Liu, D. Tran,
B. Standley, M. Aykoland, S. B. Cronin, D. Smirnov, et al.,
\textit{Nat. Phys.} \textbf{7}, 948 (2011).

\bibitem{Craciun2009} M. F. Craciun, S. Russo, M. Yamamoto, J. B. Oostinga,
A. F. Morpurgo, and S. Tarucha, \textit{Nat. Nano.} \textbf{4}, 383 (2009).

\bibitem{Koshino2009} M. Koshino and E. McCann, \textit{Phys. Rev. B} \textbf{79}, 125443(2009).

\bibitem{Russo2009} S. Russo, M. F. Craciun, M. Yamamoto, S. Tarucha, and
A. F. Morpurgo, \textit{New J. Phys.} \textbf{11}, 095018 (2009).

\bibitem{Craciun11} M. F. Craciun, S. Russo, M. Yamamoto, and S. Tarucha,
\textit{Nano Today} \textbf{6}, 42 (2011).

\bibitem{Mayo} A. S. Mayorov, R. V. Gorbachev, S. V. Morozov, L. Britnell, R. Jalil, L. A. Ponomarenko, P. Blake, K. S. Novoselov, K. Watanabe, T. Taniguchi, and A. K. Geim, \textit{Nano Lett.} \textbf{11}, 2396 (2011). 

\bibitem{Eros} H. Suzuura and T. Ando, Phys. Rev. B 65, 235412 (2002); E. Mariani and F. von Oppen, Phys. Rev. Lett. 100,
076801 (2008); E. Mariani, A. Pierce and F. von Oppen, arXiv:1110.2769.

\bibitem{Martin2010} R. T. Weitz, M. T. Allen, B. E. Feldman, J. Martin, and
A. Yacoby, \textit{Science} \textbf{330}, 812 (2010).

\bibitem{Bachtold2007} J. Moser, A. Barreiro, and A. Bachtold, \textit{Appl. Phys. Lett.} \textbf{91}, 163513 (2007).

\bibitem{Lau2008} G. Liu, J. J. Velasco, W. Bao, and C. N. Lau, \textit{Appl. Phys. Lett.} \textbf{92}, 203103 (2008).

\bibitem{Malard2009} L. M. Malard, M. H. D. G. aes, D. L. Mafra, M. S. C.
Mazzoni, and A. Jorio, \textit{Phys. Rev. B} \textbf{79}, 125426 (2009).

\bibitem{LuiHeinz2011} C. H. Lui, Z. Li, Z. Chen, P. V. Klimov, L. E. Brus, and
T. F. Heinz, \textit{Nano Lett.} \textbf{11}, 164 (2011).

\bibitem{Avouris2010} F. Xia, D. B. Farmer, Y. Lin, and P. Avouris, \textit{Nano Lett.} \textbf{10}, 715 (2010).

\bibitem{Guinea2006} F. Guinea, A. H. C. Neto, and N. M. R. Peres, \textit{Phys. Rev. B} \textbf{73}, 245426 (2006).

\bibitem{Katsnelson2011} S. Yuan, R. RoldLan, and M. I. Katsnelson, \textit{Phys. Rev. B} \textbf{84}, 125455 (2011).

\bibitem{Kumar2011} A. Kumar, W. Escoffier, J. M. Poumirol, C. Faugeras, D. P.
Arovas, M. M. Fogler, F. Guinea, S. Roche, M. Goiran, and
B. Raquet, \textit{Phys. Rev. Lett.} \textbf{107}, 126806 (2011).

\bibitem{Zaliz2011} L. Zhang, Y. Zhang, J. Camacho, M. Khodas, and I. Zaliznyak,
\textit{Nat. Phys.} \textbf{7}, 953 (2011).

\bibitem{Russo11} S. H. Jhang, M. F. Craciun, S. Schmidmeier, S. Tokumitsu, S. Russo, M. Yamamoto, Y. Skourski, J. Wosnitza, S. Tarucha, J. Eroms, and C. Strunk \textit{Phys. Rev. B} \textbf{84}, 161408 (2011).

\bibitem{Velasco} J. Velasco Jr., L. Jing, W. Bao, Y. Lee, P. Kratz, V. Aji, M. Bockrath, C.N. Lau, C. Varma, R. Stillwell, D. Smirnov, Fan Zhang, J. Jung, A.H. MacDonald, arXiv:1108.1609


\end{thebibliography}
\end{document}